\documentclass{moriond}

\def\be{\begin{equation}}
\def\ee{\end{equation}}
\def\bea{\begin{eqnarray}}
\def\eea{\end{eqnarray}}

\newcommand{\Photo}{\includegraphics[height=35mm]{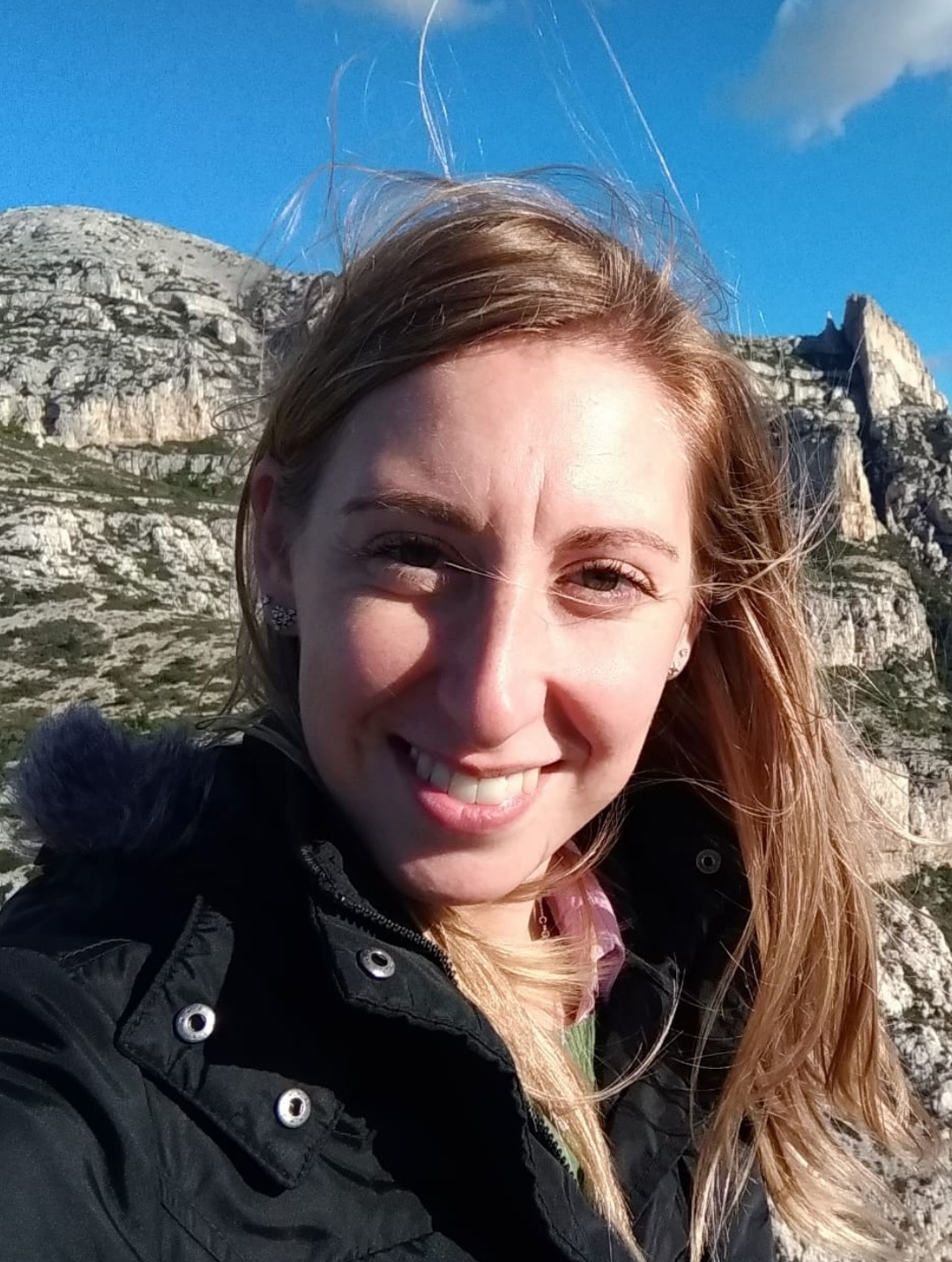}}
\newcommand{\lphi}{$\tau^- \rightarrow \ell^-\phi$}

\newcommand {\invfem}{~fb$^{-1}$}

\newcommand{\gevcc}{GeV/$c^2$}
\newcommand{\mevcc}{MeV/$c^2$}
\newcommand {\lmultau} {$L_{\mu}-L_{\tau}$}

\newcommand {\taupair} {$e^{+}e^{-}\rightarrow \tau^+ \tau^-$}
\newcommand {\epem} {$e^{+}e^{-}$}
\newcommand {\myprocess} {$e^{+}e^{-}\rightarrow \mu^+\mu^-Z'$}
\newcommand {\zprime} {$Z^\prime$}
\newcommand {\gprime} {$g^\prime$}
\newcommand {\zprimemass} {$M_{Z^\prime}$}

\newcommand {\ditaucross} {$e^{+}e^{-}\rightarrow \mu^+\mu^- \tau^+\tau^-$}
\newcommand {\taulalpha} {$e^{+}e^{-}\rightarrow \tau^+(\to \ell \alpha) \tau^- (\to \pi^+\pi^-\pi^-\nu)$}

\newcommand {\cmenergy} {$\sqrt{s}= 10.58$~GeV}
\newcommand {\bsll} {$b\to s \ell^+\ell^-$}

\usepackage{biblatex}
\usepackage{hyperref}

\begin{document}
\vspace*{4cm}
\title{Dark sectors and $\tau$ physics at Belle II}

\author{ L. Zani }

\address{INFN division of Roma Tre, Via della Vasca Navale, 84, \\Roma, 00146 Italy}

\maketitle\abstracts{The possibility of a dark sector weakly coupling to Standard Model (SM) particles through new light mediators is explored at the Belle II experiment. We present here results from three different searches, for a long-lived (pseudo)scalar particle in rare $B$ decays; for a di-tau resonance in four-muon final states, and the update on the search for a \zprime\ boson decaying invisibly. We also look for lepton flavor violation by searching for $\tau\rightarrow\ell \alpha$ decays, with $\alpha$ a new invisible boson, and we report the first untagged reconstruction of $\tau$ pairs events searching for the neutrinoless decays $\tau \to \ell \phi$.
Finally, we present the world's most precise measurement of the $\tau$ lepton mass. These studies are performed on samples from the data collected  by the Belle II detector during 2019-2021 data taking.
}

\section{Introduction}
Dark matter (DM) is one of the most compelling reasons for physics beyond the standard model (SM). Its existence has been established by several astrophysical and cosmological observations, but its origins and nature are still unknown. In recent years, especially after the null results from direct searches for heavier DM candidates, light dark sectors have become attractive, and could imply new light mediators acting as portal between SM particles and DM.
Another indisputable proof for non-SM physics is the observation of charged lepton flavor violation (LFV). Processes involving LFV can occur in the SM via weak interaction charged currents, due to neutrino oscillations, and are predicted at the level of 10$^{-50}$, which is beyond the reach of current and future experiments. 
Belle II has a unique capability to probe both light dark mediators and LFV in $\tau$ decays. Moreover, it can look for indirect signs of non-SM physics through high precision measurements of SM fundamental parameters. We report searches for new dark sectors particles, $\tau$ LFV decays and the measurement of the $\tau$ lepton mass using the data collected by the Belle II detector \cite{ref:tdr_belle} at the SuperKEKB asymmetric energy \epem\ collider. SuperKEKB mainly operates at a centre-of-mass energy of 10.58~GeV and adopts a nano-beam scheme to reach unprecedented instantaneous luminosity. 
At the time of this conference, the accelerator had achieved the peak luminosity world's record of 4.7$\times 10^{34}$~cm$^{-2}/s$ and Belle II has so far collected 424\invfem\ of data, including unique energy scan samples. It is currently in its first long shutdown.  

\section{Belle II experiment}\label{sec:exp}
The Belle II detector is a multi-purpose spectrometer  surrounding the interaction point and providing coverage of more than 90\% of the solid angle. 
The details of the Belle II detector can be found elsewhere~\cite{ref:tdr_belle}. Belle II ensures a very high reconstruction efficiency for neutral particles and excellent resolutions despite the harsh beam background environment, which are crucial when dealing with recoiling system and missing-energy final states. Additionally, it is equipped with
dedicated low-multiplicty trigger lines at hardware level, mainly based on calorimetric information, and on new topology like the single-photon or single-muon triggers, which were not available at Belle. Profiting from the well known initial state of \epem\ collisions, and its near-hermetic coverage, Belle II has a unique capability to probe signatures involving invisible final states and long-lived particles producing a displaced decay vertex. 
 Moreover, the production cross section for \taupair\ events is 0.919 nb at a c.m. energy \cmenergy, allowing Belle II to make precision measurements of $\tau$ lepton properties.
%
\section{Dark sectors results}
\begin{figure}[h!]
	\begin{center}
	\includegraphics[width=0.4\textwidth]{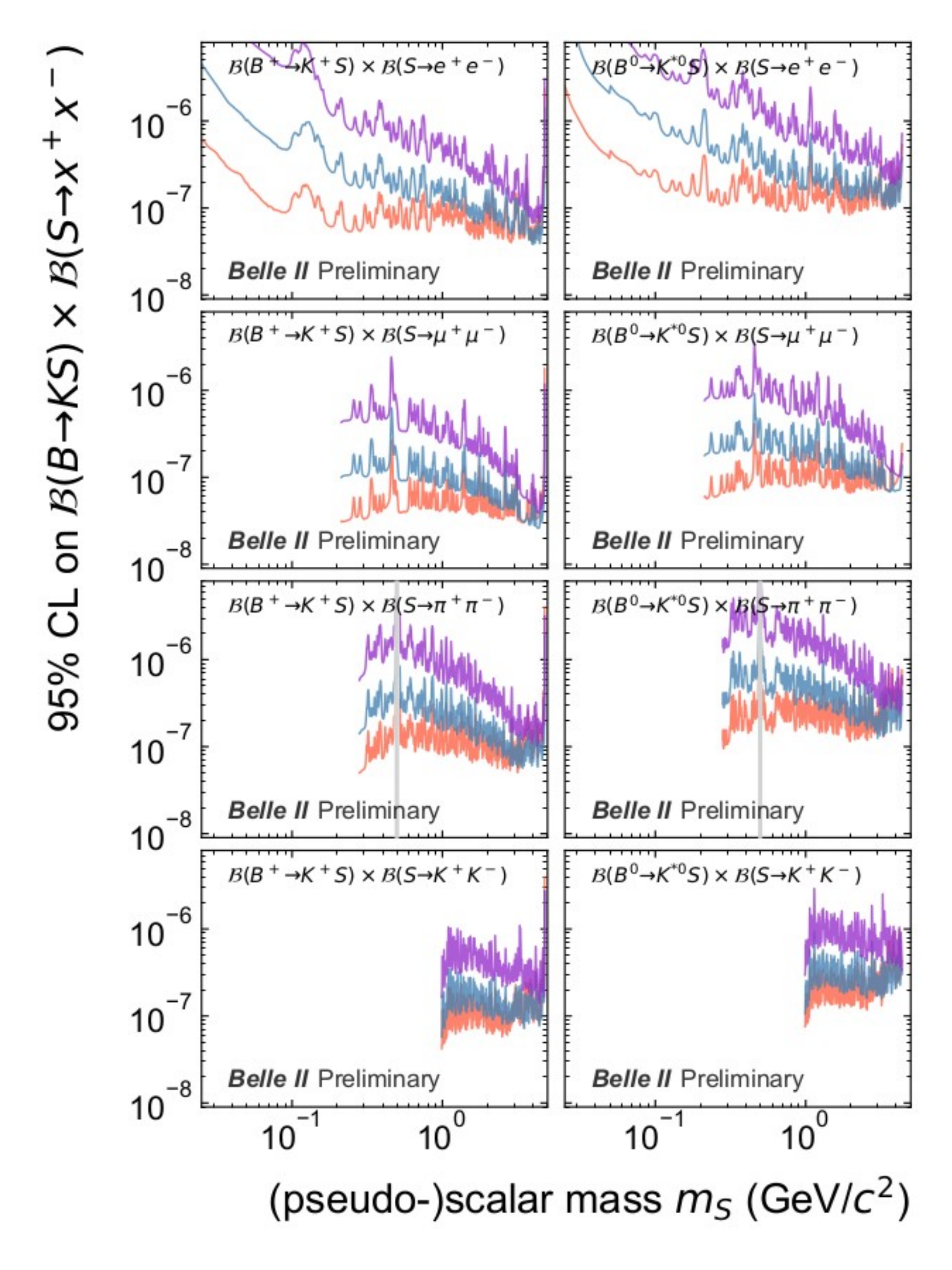}\hfill
 \includegraphics[width=0.55\textwidth]{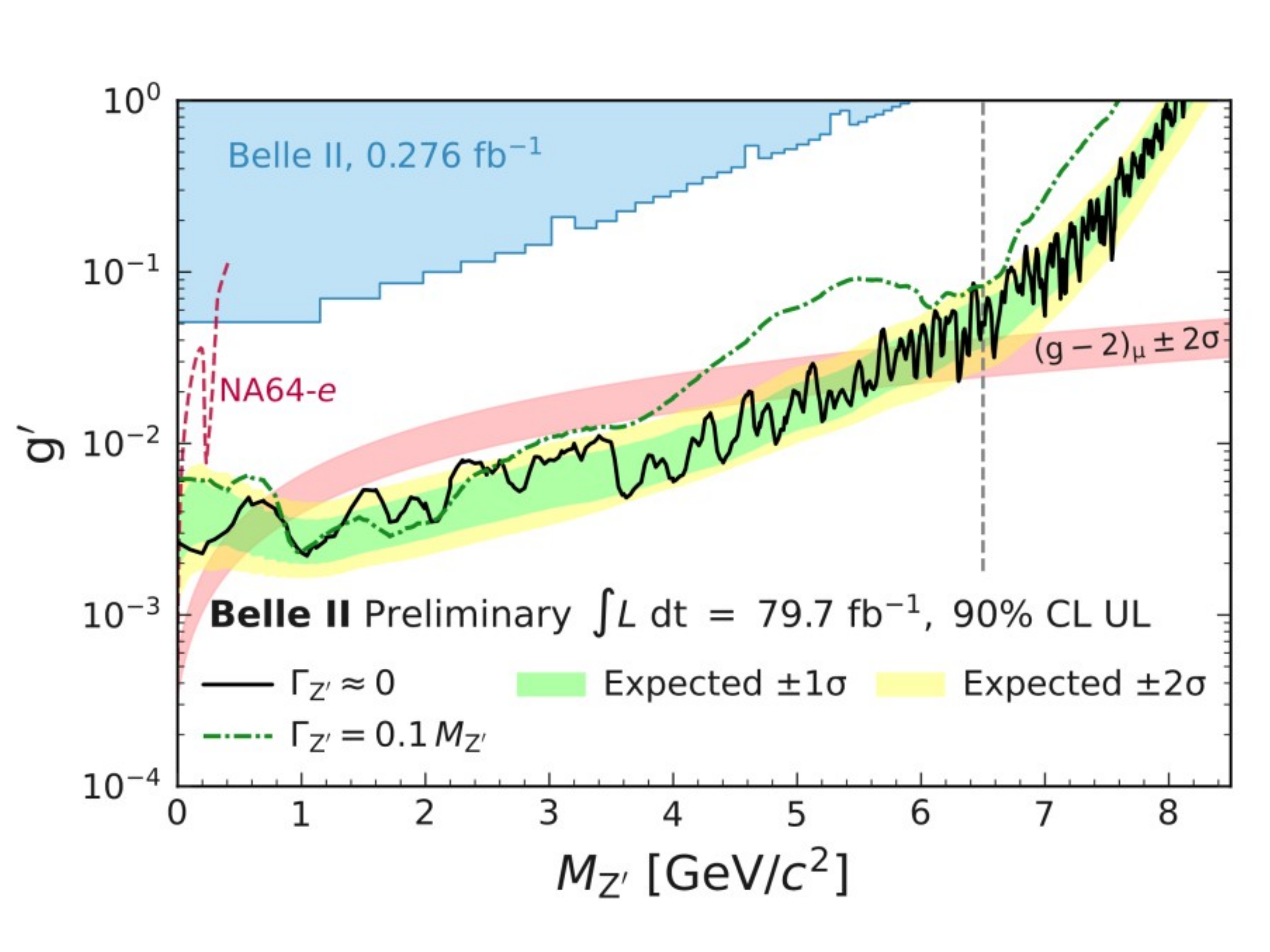}
	\caption{On the left, upper limits at 95\% CL on the product $\mathcal{B}(B\to KS)\times \mathcal{B}(S\to x^+x^-)$ as a function of the searched (pseudo)scalar mass $m_S$ are reported. On the right, 90\% CL upper limits on the coupling constant \gprime\ as a function of the searched \zprimemass\ are displayed.}
\label{fig:limits}
\end{center}
\end{figure}

Belle II can directly access the mass range that is favored by the light dark sector models, looking for new particles with masses between hundreds of MeV and a few GeV.
%
An interesting possibility is to look for new particles produced in rare $B$ meson decays involving \bsll\ transitions.
The first model-independent upper limits on long-lived (pseudo)scalar particles decaying into visible final states of two oppositely charged leptons or hadrons are set using a data set of 190\invfem\ collected during 2019-2021 data-taking period. 
We search for $B\to K(^*)S$ events selecting candidates for $S\to \ell^+\ell^-, h^+h^-$ decays, with $\ell = e, \mu$ and $h = \pi, K$, to form a displaced vertex, accompanied by a charged kaon (and additionally a pion). The combined $S$ and $K(^*)$ candidates are required to satisfy the $B$ kinematics constraints. The signal is extracted with extended maximum likelihood fits to the reduced invariant mass $m_{S} = \sqrt{M_{S\to xx}^2 -4m^2_x }$ for improving the modeling at the threshold. The only long-lived SM background is due to $K^0_S$ candidates, whose mass region is vetoed and used as control samples in data to evaluate systematic uncertainties. No significant excess is found in 190\invfem\ data and 95\% confidence level (CL) upper limits are computed on the product $\mathcal{B}(B\to KS)\times \mathcal{B}(S\to x^+x^-)$, shown as a function of the searched new particle mass $m_S$ in Figure~\ref{fig:limits}. These are the first limits set on decays to hadrons. 
%

\begin{figure}[h!]
\begin{center}
\includegraphics[width=0.48\textwidth]{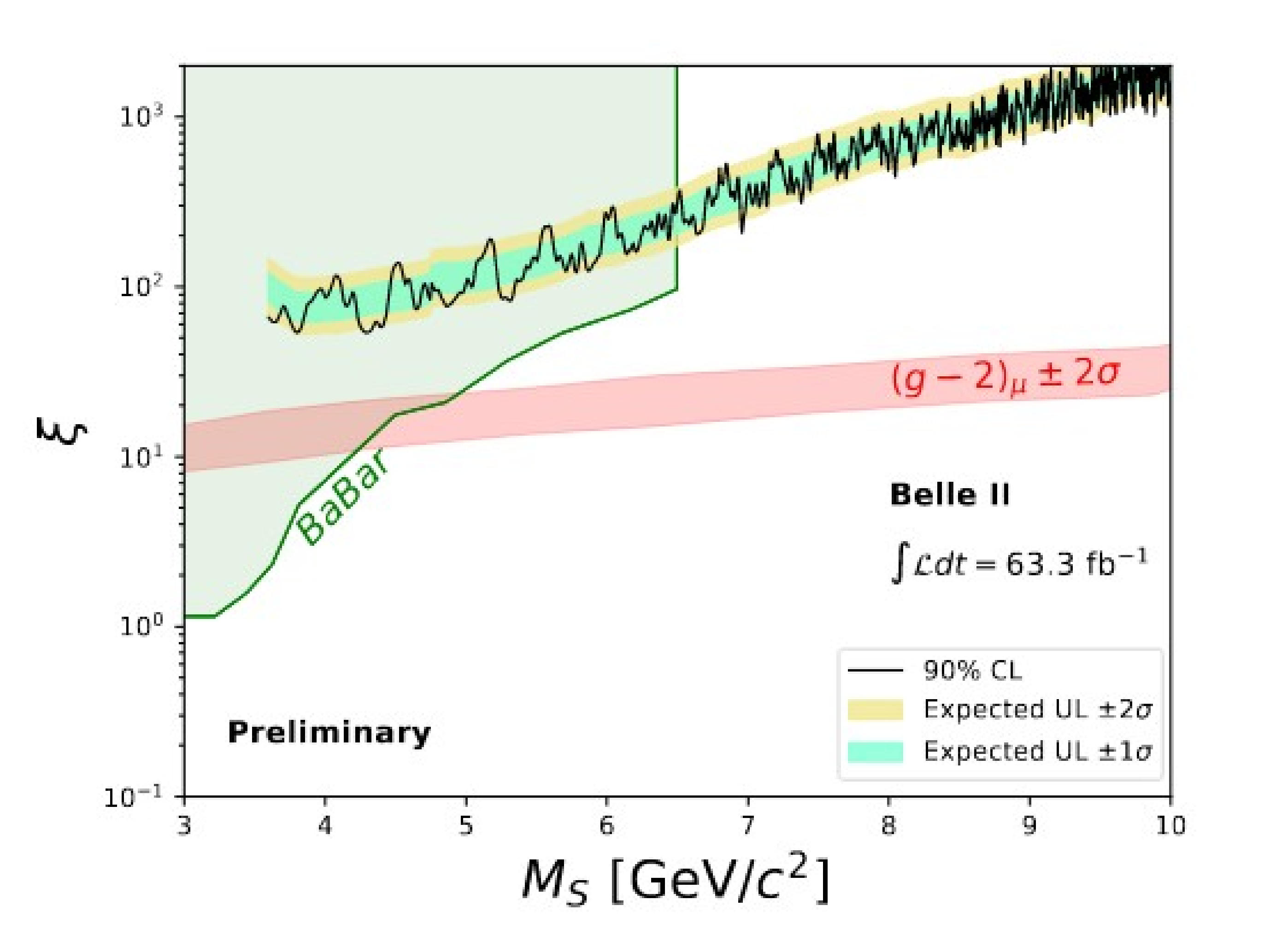}
\hfill
\includegraphics[width=0.48\textwidth]{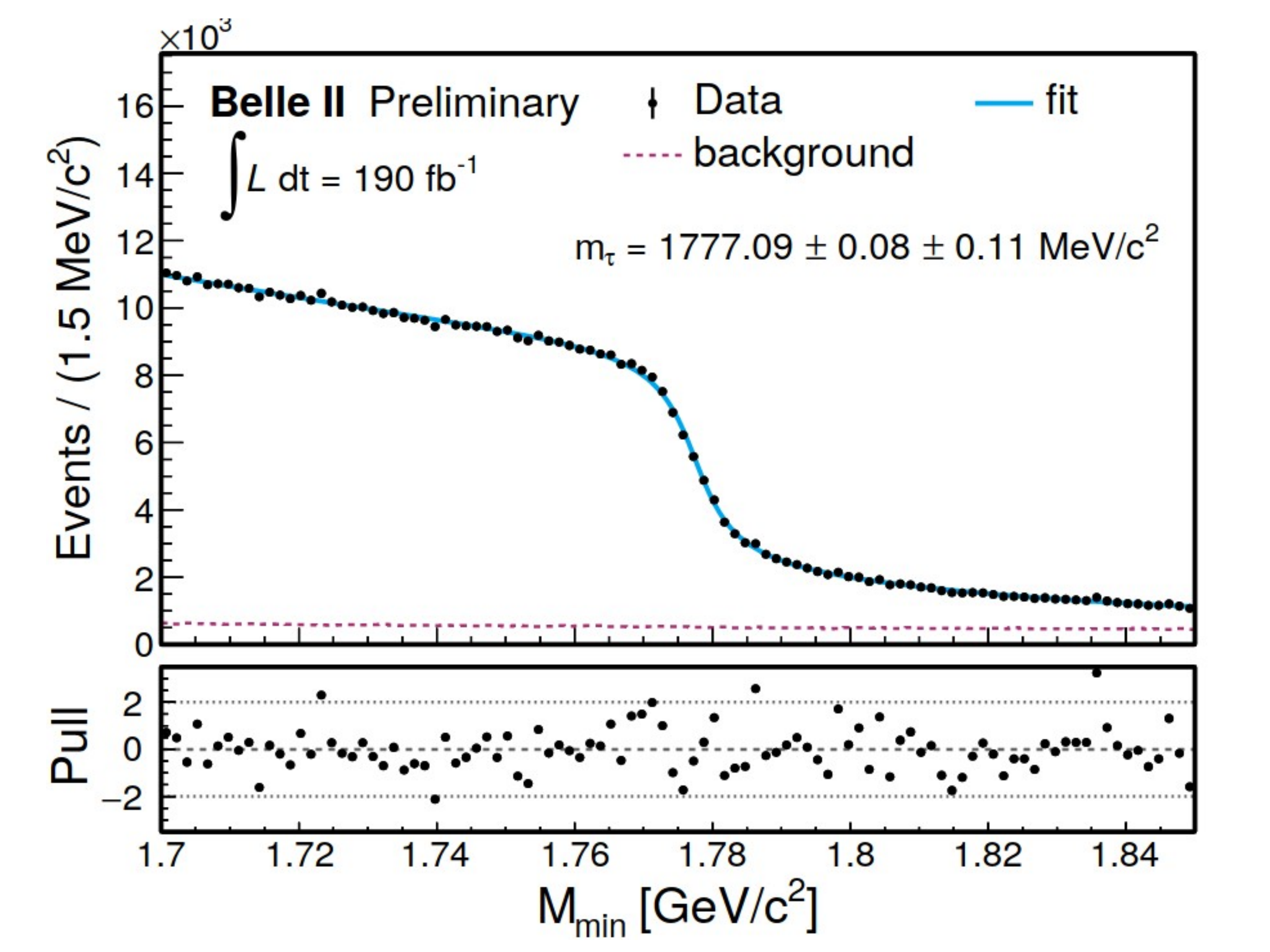}
	\caption{The 90\% CL upper limits on the leptophilic coupling constant $\xi$ as a function of the searched $M_S$ are displayed in the left plot. On the right, the pseudo-mass distribution $M_{min}$ for $\tau \to 3\pi\nu_{\tau}$ events in 190\invfem\ data, with the endpoint fit superimposed, and the measured value of the $\tau$ mass $m_{\tau}$ are shown.}
\label{fig:limits_ditau}
\end{center}
\end{figure}

A new massive vector boson that couples only to the second and third generation of leptons according to the \lmultau\ model, could explain the long-standing $(g-2)_{\mu}$ anomaly, the observed DM relic abundance and other flavor anomalies. 
We search for \myprocess\ events, where the \zprime\ is radiated off one of the muons and decays to an invisible final state with a branching fraction between 33\% and 100\%. If DM candidates are kinematically accessible, $\mathcal{B}(Z^{'} \to\chi \bar{\chi})\sim 1$. The analysis strategy is to search for a bump in the invariant mass distribution of the recoil in the centre-of-mass system against the two muons, in events where nothing else is detected.
 The main background are QED processes, namely radiative di-lepton and four-lepton final states, that mimic the signal signature of two tracks plus missing energy. 
 The kinematic properties of the signal that are specific of the production mechanism via radiation off by one muon, are fed into a neural-network trained simultaneously for all \zprime\ masses. The final signal efficiency is around 5\%, approximately constant in the whole mass range. 
 Signal yields are extracted from template fits to the recoil mass squared, in bins of recoil polar angle. We find no excess consistent with signal in 80\invfem\ of data and set 90\% CL upper limits on the cross section $\sigma(e^+e^- \to \mu^-\mu^- Z^{'}, Z' \to invisible)$. We also calculate upper limits on the coupling constant \gprime\  as a function of the \zprime\ mass, within the \lmultau\ framework and assuming $\mathcal{B}(Z^{'} \to\chi \bar{\chi})= 1$. These results, shown in the right plot in Figure~\ref{fig:limits}, exclude the region favored by the $(g-2)_{\mu}$ anomaly for the range $0.8<M_{Z'}<5$. 
 Similarly, we look for a narrow bump in the mass distribution of the recoil against two muons, in events where an additional $\tau$ pair is reconstructed in $\tau$ decays to one charged particle, $\tau \to  h, \ell \nu (\bar{\nu})$.
 Properties of the signal as final state radiation and di-tau resonant process are exploited in multi-layer perceptron neural networks to suppress the QED background, separately in eight different mass regions. The expected background is fitted directly from the recoil mass distribution in the data to avoid the impact of known mis-modeling in the simulation.
 No significant excess is found in 63\invfem\ data by fitting the recoil mass range $(3.6 < M_{recoil}(\mu\mu) < 10)$~\gevcc\ in steps of half mass resolutions, and 90\% CL upper limits are set on the quantity $\sigma($\ditaucross$)\times \mathcal{B}(X\to \tau^+\tau^-)$, which could be re-interpreted in several classes of models. This search set the world's best limits on the scalar coupling constant $\xi$ in leptophilic scalar models~\cite{ref:leptoscalar} for $M_S>6.5$~\gevcc\ ( left plot in Figure~\ref{fig:limits_ditau}).
 
\section{$\tau$ physics results}
Decays of $\tau$ leptons to new LFV bosons are postulated in many models. The process searched for in this study is \taulalpha\ and its charged conjugated, where the first $\tau$ is defined as the signal and the second one as the tag. The signal $\tau$ is searched for in its decay to a new invisible boson $\alpha$, accompanied by a lepton $\ell = e, \mu$. Its rest-frame is approximated using as energy half the collision energy $\sqrt{s}/2$ and as momentum direction the opposite to the one of the reconstructed tag $\tau$. We look for a narrow peak corresponding to the two-body decay of the signal $\tau$ in the distribution of the normalized lepton energy, over a smooth contribution coming from the irreducible background of $\tau\to \ell \nu \bar{\nu_{\ell}}$ processes. In absence of any signal excess in 63\invfem data, 95\% CL upper limits are computed on the ratio of branching fractions $\mathcal{B}(\tau \to \ell \alpha)$ normalized to  $\mathcal{B}(\tau \to\ell \nu \bar{\nu_{\ell}})$. This analysis provides limits between 2-14 times more stringent than the previous one set by ARGUS~\cite{ref:argus}. 

Possible new mediators may enhance the branching fraction for $\tau$ LFV decays \lphi\ up to observable levels of $10^{-11}-10^{-8}$, and accommodate for flavor anomalies observed in lepton flavor universality tests with $B$ decays~\cite{ref:vlq_lphi}. In contrast to previous searches for \lphi\ decays performed at Belle~\cite{ref:belle_lphi} on \taupair\ events, we apply for the first time a new \textit{untagged} approach. Only the signal $\tau$ decay to a $\phi$ meson candidate and a lepton, either muon  or electron, is explicitly reconstructed and the other $\tau$ (\textit{tag}) is not required to decay to any specific known final state. Event kinematics features and signal properties are used in a BDT classifier to suppress the background, with double the final signal efficiency for the muon mode with respect to previous analyses. Yields are extracted with a Poisson counting experiment approach from windows peaking at the known $\tau$ mass and at zero in the 2D plane of $(M_{\tau}, \Delta E_{\tau})$, respectively, with $\Delta E_{\tau}$ the difference between the reconstructed energy of the signal $\tau$ in the c.m.~frame and half the collision energy. We find no significant excess and set 90\% CL upper limits on the branching fractions to be $\mathcal{B}_{\mathrm{UL}}(\tau \to e\phi)=23\times 10^{-8}$ and $\mathcal{B}_{\mathrm{UL}}(\tau \to \mu \phi)=9.7\times 10^{-8}$.

 Lepton properties are fundamental parameters of the SM and need to be measured with the highest precision. Belle II is suitable to access several $\tau$ lepton properties. By applying the pseudo-mass $M_{min}$ technique to reconstructed \taupair\ events from 190\invfem\ data, we provide the world's most precise measurement of the $\tau$ mass $M_{\tau}$. The measured value is extracted from a fit to the endpoint of the distribution $M_{min}=\sqrt{M_{3\pi}^2+2(\sqrt{s}/2-E^*_{3\pi})(E^*_{3\pi}-P^*_{3\pi})}$, computed from events where the signal $\tau$ is reconstructed in its decays to three charged pions and the other $\tau$ decaying into one charged particle.
An excellent control of the systematic sources, dominated by the calibration of the beam energies and the charged-particle momenta scale, is required to reduce the total systematic uncertainty to 0.11~\mevcc, achieving the most precise measurement to date of the $\tau$ lepton mass of $1777.09\pm 0.08_{\mathrm{stat}}\pm 0.11_{\mathrm{sys}}$.

\printbibliography[heading=bibintoc]
\end{document}